\pdfoutput=1
\documentclass[twocolumn,superscriptaddress,aps,preprintnumbers,amsmath,amssymb,prl,nofootinbib]{revtex4}

\usepackage{amsmath}
\usepackage{amssymb}
\usepackage{amsfonts}
\usepackage{graphicx}
\usepackage{xcolor}
\usepackage{xfrac}
\usepackage{comment}
\usepackage{pifont}
\usepackage{physics}
\usepackage{fourier}
\usepackage{hyperref}
\usepackage{bm}
\usepackage{enumitem}

\definecolor{rossoferrari}{HTML}{D9073D}
\definecolor{mediumblue}{HTML}{0000CD}
\definecolor{forestgreen}{HTML}{228B22}
\definecolor{desy_blue}{HTML}{009EE2}
\definecolor{desy_orange}{HTML}{FD8800}
\definecolor{light_pink}{rgb}{1,0.4,0.4}
\definecolor{light_blue}{rgb}{0.284602,0.317763,0.963947}
\hypersetup{setpagesize=false,bookmarksnumbered=true,bookmarksopen=true,%
colorlinks=true,linkcolor=light_blue,urlcolor=rossoferrari,citecolor=rossoferrari,linktocpage=false}

\begin{document}


\preprint{MITP-25-002}
\preprint{MS-TP-25-01}

\title{A lower bound on the right-handed neutrino mass from wash-in leptogenesis}

\author{Martin A.\ Mojahed}
\email{mojahedm@uni-mainz.de}
\affiliation{Physics Department T70, Technical University of Munich, 85748 Garching, Germany}
\affiliation{PRISMA$+$ and MITP, Johannes Gutenberg University Mainz, 55099 Mainz, Germany}

\author{Kai Schmitz}
\email{kai.schmitz@uni-muenster.de}
\affiliation{Institute for Theoretical Physics, University of M\"unster, 48149 M\"unster, Germany}

\author{Dominik Wilken}
\email{dwilken2@uni-muenster.de}
\affiliation{Institute for Theoretical Physics, University of M\"unster, 48149 M\"unster, Germany}


\begin{abstract}
Leptogenesis is an attractive scenario for the generation of the baryon asymmetry of the Universe that relies on the dynamics of right-handed neutrinos (RHNs) in the seesaw extension of the Standard Model. In standard thermal leptogenesis, the RHN mass scale $M_N$ is subject to the Davidson--Ibarra bound, $M_N \gtrsim 10^9\,\textrm{GeV}$, which builds on the assumption that RHN decays are responsible for the violation of both charge-parity invariance ($CP$) and baryon-minus-lepton number ($B\!-\!L$). In this paper, we relax this assumption in the context of the more flexible framework of wash-in leptogenesis, in which $CP$ violation is encoded in the initial conditions and the only remaining task of the RHN decays is to violate $B\!-\!L$. Solving the relevant set of Boltzmann equations for vanishing initial baryon and lepton numbers (i.e., $B = L_e = L_\mu = L_\tau = 0$ initially), we find that, in wash-in leptogenesis, the RHN mass scale can be as low as 7~TeV. Wash-in leptogenesis at such low RHN masses requires the presence of a primordial charge asymmetry between right-handed electrons and left-handed positrons. We discuss several possibilities for the origin of such an asymmetry and comment on its implications for the chiral  instability of the Standard Model plasma.
\end{abstract}


\date{\today}
\maketitle


\noindent
\textbf{Introduction}\,---\,The baryon asymmetry of the Universe (BAU), commonly quantified in terms of the baryon-to-photon ratio $\eta_B^{\rm obs} = n_B^0/n_\gamma^0 \simeq 6.12 \times 10^{-10}$~\cite{Planck:2018vyg,ParticleDataGroup:2024cfk}, cannot be explained by the Standard Model (SM) of particle physics and hence provides evidence for new physics beyond the SM (BSM)~\cite{Sakharov:1967dj}. An attractive BSM scenario capable of dynamically generating the BAU in the early Universe consists of baryogenesis via leptogenesis~\cite{Fukugita:1986hr}, which relies on the interactions of right-handed neutrinos (RHNs) in the seesaw extension of the SM~\cite{Minkowski:1977sc,Yanagida:1979as,Yanagida:1980xy,Gell-Mann:1979vob,Mohapatra:1979ia}, thus linking the generation of the BAU to the origin of the tiny masses of the left-handed neutrinos in the SM~\cite{Buchmuller:2005eh,Chun:2017spz,Bodeker:2020ghk}. In its standard implementation, leptogenesis proceeds via the out-of-equilibrium decays of heavy RHNs, which simultaneously violate charge--parity invariance ($CP$) and baryon-minus-lepton number ($B\!-\!L$). These decays first produce a primordial lepton asymmetry, which is subsequently converted by means of the chemical transport in the thermal plasma, including electroweak sphalerons~\cite{Kuzmin:1985mm}, to a primordial baryon asymmetry.

The requirement of sufficient $CP$ violation in RHN decays bounds the RHN mass scale from below, $M_N \gtrsim 10^9\,\textrm{GeV}$. This constraint is known as the Davidson--Ibarra bound~\cite{Davidson:2002qv,Buchmuller:2002rq} (see Ref.~\cite{Garbrecht:2024xfs} for a recent refinement). It assumes a hierarchical RHN mass spectrum (e.g., $M_3 \gg M_2 \gg M_1 \equiv M_N$ in models with three RHNs) and can be circumvented in the case of a quasi-degenerate RHN mass spectrum (e.g., $M_3 \gg M_2 \simeq M_1$ or $M_3 \simeq M_2 \simeq M_1$), which gives rise to the scenario of resonant leptogenesis~\cite{Pilaftsis:2003gt,Pilaftsis:2005rv} (see also Refs.~\cite{Klaric:2020phc,Klaric:2021cpi}). In this paper, we will, however, ignore the possibility of resonantly enhanced $CP$ asymmetry in RHN decays and focus instead on an alternative route to low-scale leptogenesis at RHN masses $M_N \ll 10^9\,\textrm{GeV}$: wash-in leptogenesis~\cite{Domcke:2020quw}. As we will show in this paper, wash-in leptogenesis can still successfully operate at RHN masses slightly below $10\,\textrm{TeV}$. In alternative scenarios, such as resonant leptogenesis or freeze-in leptogenesis via RHN oscillations (Akhmedov--Rubakov--Smirnov or ARS leptogenesis~\cite{Akhmedov:1998qx}), it is possible to produce the baryon asymmetry at even lower RHN masses, partially even in the sub-GeV regime~\cite{Klaric:2021cpi}. We emphasize that wash-in leptogenesis differs from these scenarios: unlike resonant leptogenesis, it requires no approximate mass degeneracy among the RHN mass states; and unlike both resonant and ARS leptogenesis, it does not depend on the amount of $CP$ violation in the RHN sector [see the discussion below Eq.~\eqref{eq:BEqDeltalpha} for more details]. 

Indeed, the key idea behind wash-in leptogenesis is to liberate the RHN decays from the requirement of sufficient $CP$ violation. In wash-in leptogenesis, RHN interactions are treated as mere spectator processes and on the same footing as the electroweak sphalerons. From the perspective of wash-in leptogenesis, sphalerons and RHN decays represent nothing but sources of $B\!+\!L$ violation and $B\!-\!L$ violation, respectively. $CP$ violation, on the other hand, is attributed to new $CP$-violating dynamics at high temperatures, $T \gg M_N$, which set the required $CP$-violating initial conditions for wash-in leptogenesis. In this sense, wash-in leptogenesis does not represent a complete baryogenesis scenario on its own, but calls for an ultraviolet completion\,---\,a chargegenesis mechanism that is responsible for the generation of primordial $CP$-violating charge asymmetries at high energies. Wash-in leptogenesis then acts on the nontrivial configuration of chemical potentials induced by chargegenesis and reshuffles the primordial charge asymmetries in a $B\!-\!L$-violating manner. The SM notably offers more than ten conserved global charges at high temperatures that could be produced during chargegenesis~\cite{Domcke:2020quw}, which renders baryogenesis via leptogenesis via chargegenesis significantly more versatile than baryogenesis via leptogenesis. As a consequence, the range of possible chargegenesis scenarios is extremely broad; primordial charge asymmetries suitable for wash-in leptogenesis can, e.g., be produced during axion inflation~\cite{Domcke:2022kfs}, by evaporating primordial black holes~\cite{Schmitz:2023pfy}, in the decay of heavy Higgs doublets~\cite{Mukaida:2024eqi}, or via new nonminimal gravitational interactions~\cite{Mojahed:2024yus}.

As the temperature of the thermal bath drops, the last global charge to be erased by the SM interactions is the asymmetry between right-handed electrons and left-handed positrons, $q_e = n_e - \bar{n}_e$. Therefore, as long as the temperature scale of wash-in leptogenesis exceeds the equilibration temperature of the electron Yukawa interaction, $T_e \simeq 85\,\textrm{TeV}$~\cite{Bodeker:2019ajh}, it is always possible to realize successful wash-in leptogenesis by assuming a nonzero primordial electron charge, $q_e^{\rm ini} \neq 0$. Correspondingly, the authors of Ref.~\cite{Domcke:2020quw} concluded that wash-in leptogenesis can take place at temperatures as low as $T \simeq 100\,\textrm{TeV}$. The purpose of the present paper is to make this statement more precise and study wash-in leptogenesis at $T \lesssim T_e$, where $q_e/T^3$ can no longer be treated as a constant, i.e., in a temperature regime where the asymmetry wash-in by RHN processes competes with the wash-out of the primordial $q_e$ input charge due to the progressing equilibration of the electron Yukawa interaction. To this end, we shall extend the treatment in Ref.~\cite{Domcke:2020quw} and supplement the Boltzmann equations for the three flavor asymmetries $q_{\Delta_\alpha}$, where $\Delta_\alpha = B/3 - L_\alpha$ and $\alpha = e,\mu,\tau$, that are being produced during wash-in leptogenesis by a Boltzmann equation for the time-dependent electron charge $q_e$. Solving this coupled set of Boltzmann equations will then allow us to conclude that wash-in leptogenesis is able to produce the BAU for RHN masses as low as $M_N \simeq 7\,\textrm{TeV}$. In summary, in standard thermal leptogenesis where RHN decays act as a source of both $CP$ and $B\!-\!L$ violation, the RHN mass scale must satisfy the Davidson--Ibarra bound, $M_N \gtrsim 10^9\,\textrm{GeV}$; if RHN decays only act as a source of $B\!-\!L$ violation, it suffices to demand that $M_N \gtrsim 7\,\textrm{TeV}$.


\medskip\noindent
\textbf{Boltzmann equations}\,---\,Let us begin by recalling a few key results from Ref.~\cite{Domcke:2020quw}. The evolution of the $q_{\Delta_\alpha}$ flavor asymmetries during $N_1$ leptogenesis is governed by~\cite{Pilaftsis:2003gt,Pilaftsis:2005rv}
\begin{equation}
\label{eq:BEqDeltalpha}
-\left(\partial_t + 3 H\right) q_{\Delta_\alpha} = \varepsilon_{1\alpha}\, \Gamma_1 \left(n_{N_1}-n_{N_1}^{\rm eq}\right) - \sum_\beta \gamma_{\alpha\beta}^{\rm w}\,\frac{\mu_{\ell_\beta}+\mu_\phi}{T} \,.
\end{equation}
Here, the negative sign on the left-hand side reflects the fact that the lepton flavor number $L_\alpha$ enters $\Delta_\alpha$ with a negative sign. The two terms on the right-hand side correspond to the standard source term for $q_{\Delta_\alpha}$ and the standard wash-out term in thermal leptogenesis, respectively. Here, the total wash-out interaction density receives contributions from RHN inverse decays as well as from lepton-number violating ($\Delta L = 2$) and lepton-flavor-violating ($\Delta L = 0$) processes. This interaction density is multiplied by $\mu_{\ell_\beta}+\mu_\phi$, i.e., the sum of the chemical potentials of the SM charged-lepton and Higgs doublets. At $T \sim T_e$, the chemical potentials of the 16 SM particle multiplets, $\mu_i$ ($i = e, \mu, \tau, \ell_e, \ell_\mu, \ell_\tau, u, c, t, d, s, b, Q_1, Q_2, Q_3, \phi$), are subject to 16 constraints: 11 conditions for 11 linearly independent Yukawa and sphaleron interactions in chemical equilibrium and 5 conservation laws for 5 conserved global charges ($\Delta_{e,\mu,\tau}$, hypercharge $Y$, and the asymmetry in right-handed electrons). Solving this linear system of constraint equations for the 16 chemical potentials $\mu_i$, one finds~\cite{Domcke:2020kcp,Domcke:2020quw}
\begin{equation}
\begin{pmatrix}
		\mu_{\ell_e}    + \mu_\phi \\
		\mu_{\ell_\mu}  + \mu_\phi \\
		\mu_{\ell_\tau} + \mu_\phi
	\end{pmatrix}
	= 
	\begin{pmatrix}
		-\frac{5}{13} \\
		 \frac{4}{37} \\
		 \frac{4}{37}
	\end{pmatrix} \,\bar{\mu}_e 
	-
	\begin{pmatrix}
		\frac{6}{13} &  0              & 0             \\
		0            &  \frac{41}{111} & \frac{4}{111} \\
		0            &  \frac{4}{111}  & \frac{41}{111} 
	\end{pmatrix}
	\begin{pmatrix}
		\bar{\mu}_{\Delta_e}\\
		\bar{\mu}_{\Delta_\mu}\\
		\bar{\mu}_{\Delta_\tau}
	\end{pmatrix} \,,
\end{equation}
or in more compact notation,
\begin{equation}
\label{eq:mulmuphi}
\mu_{\ell_\alpha}    + \mu_\phi = \bar{\mu}_\alpha^0 - \sum_\beta C_{\alpha\beta}\,\bar{\mu}_{\Delta_\beta} \,.
\end{equation}
where the second term on the right-hand side contains the standard flavor coupling matrix $C_{\alpha\beta}$~\cite{Barbieri:1999ma,Abada:2006fw,Nardi:2006fx,Abada:2006ea,Blanchet:2006be,Antusch:2006cw} induced by the spectator processes in the thermal bath~\cite{Buchmuller:2001sr,Garbrecht:2014kda,Garbrecht:2019zaa}, while the first term is the relevant term for wash-in leptogenesis. 

Inserting the affine relation in Eq.~\eqref{eq:mulmuphi} into Eq.~\eqref{eq:BEqDeltalpha}, the Boltzmann equation for the three lepton flavor asymmetries can be brought into the following form,
\begin{align}
\label{eq:BEq2}
& -\left(\partial_t + 3 H\right) q_{\Delta_\alpha} = \\ & \varepsilon_{1\alpha}\, \Gamma_1 \left(n_{N_1}-n_{N_1}^{\rm eq}\right) 
- \sum_\beta \Gamma_{\alpha\beta}^{\rm w} \left(q_\beta^0 - \sum_\sigma C_{\beta\sigma}\, q_{\Delta_\sigma}\right) \,, \nonumber
\end{align}
where $q_\alpha^0 \equiv \sfrac{1}{6}\:\bar{\mu}_\alpha^0 T^2$, $q_{\Delta_\alpha} \equiv \sfrac{1}{6}\:\bar{\mu}_{\Delta_\alpha}T^2$, and $\Gamma_{\alpha\beta} = 6/T^3\,\gamma_{\alpha\beta}$. This evolution equation is linear in the $q_{\Delta_\alpha}$, which allows us to write $q_{\Delta_\alpha} = q_{\Delta_\alpha}^{\rm th} + q_{\Delta_\alpha}^{\rm win}$ and split Eq.~\eqref{eq:BEq2} into two equations,
\begin{align}
-\left(\partial_t + 3 H\right) q_{\Delta_\alpha}^{\rm th} & = \varepsilon_{1\alpha}\, \Gamma_1 \left(n_{N_1}-n_{N_1}^{\rm eq}\right) - \sum_{\beta,\sigma} \Gamma_{\alpha\beta}^{\rm w} C_{\beta\sigma}\, q_{\Delta_\sigma}^{\rm th} \,, \label{eq:qth} \\
-\left(\partial_t + 3 H\right) q_{\Delta_\alpha}^{\rm win} & =  - \sum_\beta \Gamma_{\alpha\beta}^{\rm w} \left(q_\beta^0 - \sum_\sigma C_{\beta\sigma}\, q_{\Delta_\sigma}^{\rm win}\right) \,.
\label{eq:qwin}
\end{align}
Indeed, the sum of these two equations returns again the original Boltzmann equation in Eq.~\eqref{eq:BEq2}. The first equation, Eq.~\eqref{eq:qth}, is identical to the standard Boltzmann equation for thermal leptogenesis (possibly in the resonant regime) in a trivial chemical background (i.e., no affine $q_\alpha^0$ term) and fully accounts for the effect of $CP$ violation in the RHN sector. The second equation, Eq.~\eqref{eq:qwin}, on the other hand, describes the evolution of $q_{\Delta_\alpha}^{\rm win}$, i.e., the contributions to the lepton flavor asymmetries produced during wash-in leptogenesis. By definition, Eq.~\eqref{eq:qwin} is completely independent of the RHN $CP$ violation parameters $\varepsilon_{1\alpha}$. In the following, we shall focus on the wash-in contributions to the total lepton flavor asymmetries and assume that $q_{\Delta_\alpha}^{\rm th} \ll  q_{\Delta_\alpha}^{\rm win}$ for all $\alpha$. In other words, we shall work in a part of the RHN parameter space where the RHN $CP$ violation parameters $\varepsilon_{1\alpha}$ are negligibly small and neither resonant nor ARS leptogenesis manage to explain the observed value of the BAU. Instead, we will assume a nonvanishing primordial charge asymmetry, $\bar{\mu}_e \neq 0$, which significantly changes the character of the processes contributing to the rate $\Gamma_{\alpha\beta}^{\rm w}$ in Eq.~\eqref{eq:qwin}: Typically, these processes describe \textit{wash-out} effects that drive the $q_{\Delta_\alpha}$ asymmetries to zero; see Eq.~\eqref{eq:qth}. However, in the presence of a primordial input charge $q_e$, the wash-out processes begin to \textit{wash in} nonzero lepton flavor asymmetries proportional to $q_e$, even if $q_{\Delta_\alpha} = 0$ initially. We notably attribute the origin of the primordial charge asymmetry $q_e$ to new $CP$-violating dynamics in a different sector (see the discussion above), such that every single term in Eq.~\eqref{eq:qwin} is insensitive to the effect of $CP$ violation in the RHN sector.

Assuming a constant comoving charge density, $q_e/T^3 = \textrm{const}$, wash-in leptogenesis in the strong wash-in regime (conventionally, strong wash-out regime) at $T \gtrsim T_e$ results in a $B\!-\!L$ asymmetry $q_{B-L} = -\sfrac{3}{10}\,q_e$~\cite{Domcke:2020quw}. Around $T \sim  T_e$, the assumption of constant $q_e/T^3$, however, begins to break down and $q_e/T^3$ needs to be described as a time-dependent quantity. The relevant Boltzmann equation for $q_e$ around the equilibration temperature of the electron Yukawa interaction has been derived in Ref.~\cite{Bodeker:2019ajh},
\begin{equation}
\label{eq:BEqe}
\left(\partial_t + 3 H\right) q_e = - \Gamma_e\left[\frac{711}{481}\,q_e + \frac{5}{13}\,q_{\Delta_e} - \frac{4}{37}\left(q_{\Delta_\mu} + q_{\Delta_\tau}\right) \right] \,,
\end{equation}
where $\Gamma_e$ is the equilibration rate of right-handed electrons.

Let us now simplify the notation and bring the Boltzmann equations in Eqs.~\eqref{eq:BEqDeltalpha} and \eqref{eq:BEqe} into a more convenient form. First, we note that, around $T \sim T_e$, the standard wash-out rate is dominated by RHN inverse decays, such that
\begin{equation}
\gamma_{\alpha\beta}^{\rm w} = p_\alpha\,\delta_{\alpha\beta}\,\frac{T^3}{6}\,\Gamma_{\rm w} \,, \qquad p_\alpha = \frac{\Gamma_{1\alpha}}{\Gamma_1} \,.
\end{equation}
$p_\alpha$ is the branching ratio for RHN decays into lepton flavor $\alpha$, and $\Gamma_{\rm w}$ is the rate of wash-out via inverse decays~\cite{Buchmuller:2004nz},
\begin{equation}
\Gamma_{\rm w} = \frac{3g_N}{\pi^2}\,\frac{\widetilde{m}_1M^2}{8\pi v^2}\,z^2 K_1\left(z\right) \,.
\end{equation}
Here, $g_N = 2$ counts the RHN spin degrees of freedom (DOFs), $M$ denotes the RHN mass, $z = M/T$ functions as a time variable, $K_1$ is a modified Bessel function, $v = 174\,\textrm{GeV}$ is the Higgs vacuum expectation value, and $\widetilde{m}_1$ is an effective light neutrino mass defined in terms of $M$ and the neutrino Dirac mass matrix $m_D$ in the seesaw model~\cite{Plumacher:1996kc},
\begin{equation}
\widetilde{m}_1 = \frac{\left(m_D^\dagger m_D^{\vphantom{\dagger}}\right)_{11}}{M_1} \,.
\end{equation}



\begin{figure*}%
    \centering
    {{\includegraphics[width=0.49\linewidth]{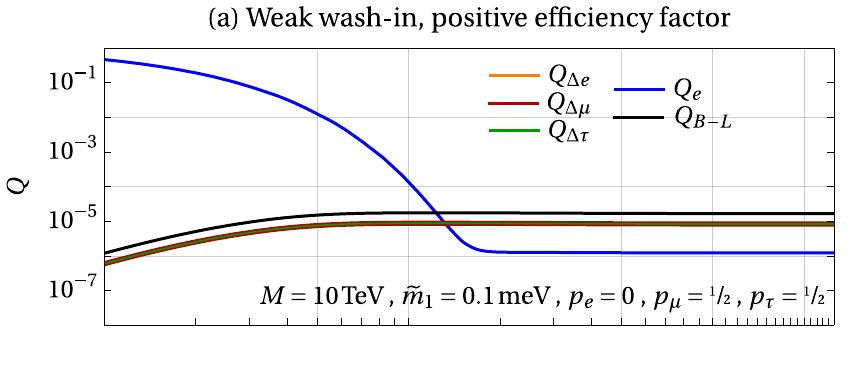}}}%
    \quad
    {{\includegraphics[width=0.49\linewidth]{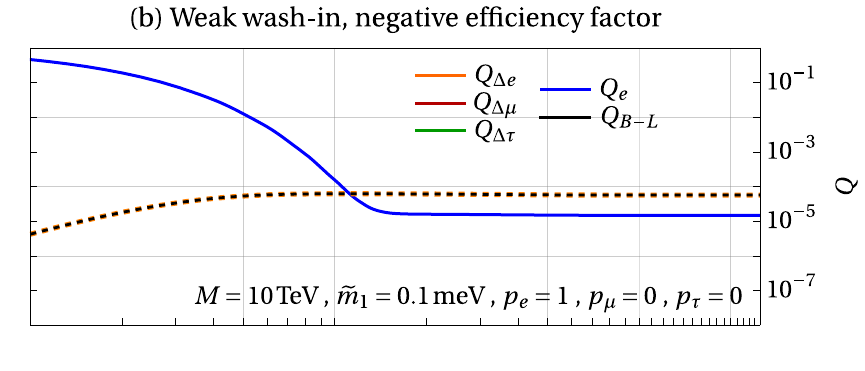} }}%
    \quad
    {{\includegraphics[width=0.49\linewidth]{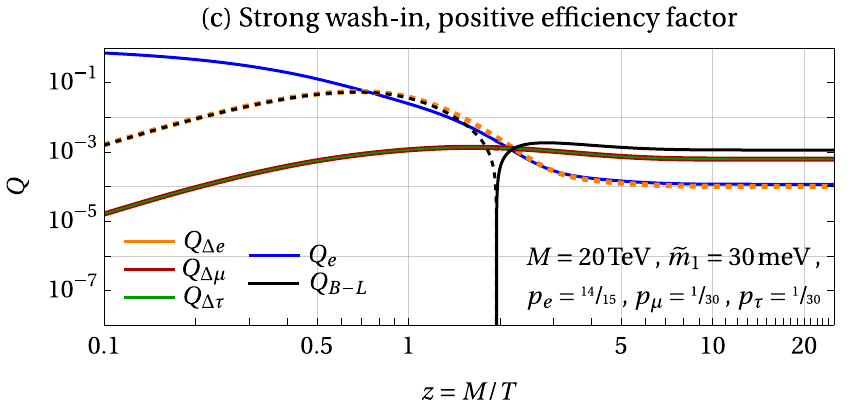} }}%
    \quad
    {{\includegraphics[width=0.49\linewidth]{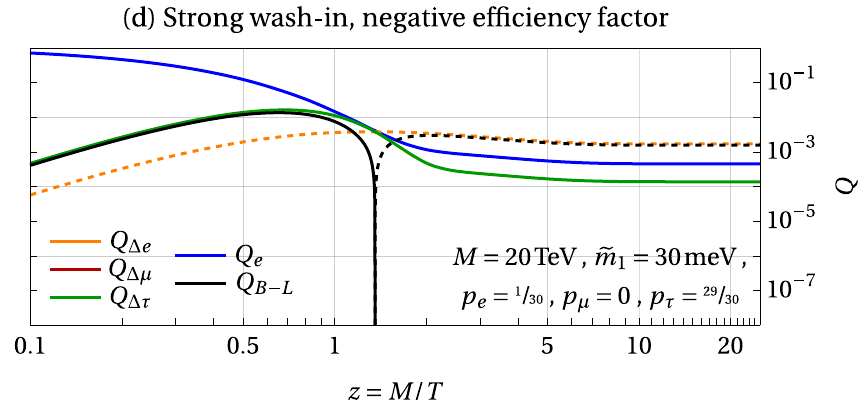} }}%
    \caption{Example solutions of the set of Boltzmann equations in Eqs.~\eqref{eq:BE1} to \eqref{eq:BE4} in the weak\,/\,strong wash-in regime (top\,/\,bottom row). Solid\,/\,dashed lines denote positive\,/\,negative comoving asymmetries $Q$. The $Q_{B-L}$ values at $z \gg 1$ allow one to read off the efficiency factor defined in Eq.~\eqref{eq:kappa}, $Q_{B-L}\left(z\gg 1\right) = C_{\rm win}\,\kappa$. The solutions in the first\,/\,second column correspond to positive\,/\,negative $\kappa$ values. In all four plots, the RHN branching ratios $p_{e,\mu,\tau}$ are set to the optimal values that maximize $\left|\kappa\right|$ for the given neutrino masses $\widetilde{m}_1$ and $M$.}
    \label{fig:BEsols}
\end{figure*}


Meanwhile, the electron equilibration rate can be written as $\Gamma_e = C_e\,T$, where the coefficient $C_e$ receives contributions from $2\rightarrow 2$ processes involving one electron Yukawa vertex as well as from $1 \leftrightarrow 2$ decays and inverse decays of the SM Higgs bosons, right-handed electrons and left-handed leptons. In fact, the latter processes also receive corrections from $1n \leftrightarrow 2n$ scatterings with soft gauge boson exchanges, which can be taken care of by Landau--Pomeranchuk--Migdal (LPM) resummation~\cite{Landau:1953um,Landau:1953gr,Migdal:1956tc}. Taking all these processes into account, the authors of Ref.~\cite{Bodeker:2019ajh} find
\begin{align}
\label{eq:Ce}
C_e = \frac{3h_e^2}{1024\pi}\Bigg[ & h_t^2 c_t + \left(3g^2 + g'^2\right)\left(c_\ell + \ln\frac{1}{3g^2 + g'^2}\right)  \\
& + 4g'^2\left(c_e + \ln\frac{1}{4g'^2}\right) \Bigg] \,,
\end{align}
where $c_t$, $c_\ell$, and $c_e$ receive contributions from both $2\rightarrow 2$ processes and LPM-resummed $1n \leftrightarrow 2n$ processes,
\begin{align}
c_t    & = 2.82 + 1.48  \,, \\
c_\ell & = 3.52 + 0.776 \,, \\
c_e    & = 2.69 + 2.03  \,.
\end{align}
Furthermore, $h_e$ and $h_t$ in Eq.~\eqref{eq:Ce} denote the electron and top-quark Yukawa couplings, while $g$ and $g'$ are the $SU(2)_L$ and $U(1)_Y$ gauge couplings. In our numerical analysis, we start from the SM values of all four couplings at the $Z$ pole and then use their one-loop renormalization group equations (RGEs) to evolve them up to the RGE scale $\mu = \pi T$.

Next, we introduce the comoving charge asymmetries in units of the initial electron asymmetry $q_e^{\rm ini}$ at $T_{\rm ini} \gg T_e$,
\begin{equation}
Q_{\Delta_\alpha} = \left(\frac{a}{a_{\rm ini}}\right)^3\frac{q_{\Delta_\alpha}}{q_e^{\rm ini}} \,, \qquad Q_e = \left(\frac{a}{a_{\rm ini}}\right)^3\frac{q_e}{q_e^{\rm ini}} \,,
\end{equation}
where $a \propto z$ is the Friedmann--Lema\^itre--Robertson--Walker scale factor. Using our above results for the interaction rates $\Gamma_{\rm w}$ and $\Gamma_e$, we can then rewrite the Boltzmann equations in Eqs.~\eqref{eq:BEqDeltalpha} and \eqref{eq:BEqe} as a set of equations for $Q_{\Delta_\alpha}$ and $Q_e$,
\begin{align}
\label{eq:BE1}
Q'_{\Delta_e} & = \, p_e W \left[-\frac{5}{13}\,Q_e - \frac{6}{13}\,Q_{\Delta_e}\right] \,, \\
\label{eq:BE2}
Q'_{\Delta_\mu} & = p_\mu W \left[\frac{4}{37}\,Q_e - \frac{41}{111}\,Q_{\Delta_\mu} - \frac{4}{111}\,Q_{\Delta_\tau}\right] \,, \\
\label{eq:BE3}
Q'_{\Delta_\tau} & = \, p_\tau W \left[\frac{4}{37}\,Q_e - \frac{4}{111}\,Q_{\Delta_\mu} - \frac{41}{111}\,Q_{\Delta_\tau}\right] \,, \\
\label{eq:BE4}
Q_e' & = \quad - E \left[\frac{711}{481}\,Q_e + \frac{5}{13}\,Q_{\Delta_e} - \frac{4}{37}\left(Q_{\Delta_\mu} + Q_{\Delta_\tau}\right)\right] \,.
\end{align}
Here, $W$ denotes the dimensionless wash-out rate,
\begin{equation}
\label{eq:W}
W = \frac{\Gamma_{\rm w}}{zH} = \frac{6}{\pi^2}\,\frac{\widetilde{m}_1}{m_*}\,z^3K_1\left(z\right) \,,
\end{equation}
while $E$ is the dimensionless electron equilibration rate,
\begin{equation}
E = \frac{\Gamma_e}{zH} = C_e\,\frac{M_*}{M} \,.
\end{equation}
In both expressions, $H$ is the Hubble rate, which can be written as follows during the era of radiation domination,
\begin{equation}
H = \frac{T^2}{M_*} = \frac{M^2}{z^2 M_*} \,,
\end{equation}
where $M_*$ denotes the reduced Planck mass $M_{\rm Pl}$ weighted by a factor taking into account the effective number of relativistic DOFs in the SM at high temperatures, $g_* = \sfrac{427}{4}$,
\begin{equation}
M_* = \left(\frac{90}{\pi^2 g_*}\right)^{1/2} M_{\rm Pl} \simeq 7.12 \times 10^{17}\,\textrm{GeV} \,.
\end{equation}
The mass $m_*$ in Eq.~\eqref{eq:W}, on the other hand, is defined as~\cite{Buchmuller:2004nz}
\begin{equation}
m_* = \frac{8\pi v^2}{M_*} \simeq 1.07\,\textrm{meV} \,.
\end{equation}

These definitions conclude our derivation of the Boltzmann equations. As evident from Eqs.~\eqref{eq:BE1} to \eqref{eq:BE4}, wash-in leptogenesis at $T \sim T_e$ effectively depends on four parameters: (1) the effective neutrino mass $\widetilde{m}_1$, which controls the strength of asymmetry wash-in, (2) the RHN mass $M$, which determines the temperature scale of wash-in leptogenesis and hence the importance of $q_e$ wash-out, and (3 and 4) the branching ratios $p_e$ and $p_\mu$ ($p_\tau$ follows from  $p_e + p_\mu + p_\tau = 1$).


\medskip\noindent
\textbf{Numerical results}\,---\,We are now ready to numerically solve the Boltzmann equations for $Q_{\Delta_\alpha}$ and $Q_e$. In doing so, we shall assume vanishing initial lepton flavor asymmetries and a nonvanishing primordial electron asymmetry (stemming from an unspecified chargegenesis mechanism),
\begin{equation}
Q_{\Delta_\alpha} \left(z\ll 1\right) = 0 \,, \qquad Q_e \left(z\ll 1\right) = 1 \,.
\end{equation}
These initial conditions define the minimal scenario that we are interested in in this paper, in which the RHN interactions during wash-in leptogenesis are the only source of lepton number and lepton flavor violation. In particular, the first condition corresponds to the standard starting point of other typical leptogenesis scenarios in the literature: before the dynamics in the RHN sector become relevant, we set $Q_{\Delta_\alpha} = 0$ for all $\alpha$. This assumption imposes restrictions on the chargegenesis mechanism setting the initial conditions for wash-in leptogenesis. Our analysis in this paper only applies to scenarios of chargegenesis that result in $Q_e \neq 0$, possibly other charges, but $Q_{\Delta_\alpha} = 0$ for all $\alpha$. Chargegenesis may, e.g., also lead to nonzero $Q_\mu \neq Q_e$ and nonzero $Q_\tau \neq Q_e$, meaning that flavor universality may be violated during chargegenesis. Any nonzero primordial muon or tau charge asymmetries will simply be washed out by the respective Yukawa interactions before the temperature regime relevant for our analysis in this paper is reached. This is explicitly illustrated by Eqs.~(4.6) and (4.8)--(4.11) in Ref.~\cite{Domcke:2022kfs}, which are general and go beyond the specific scenario discussed in Ref.~\cite{Domcke:2022kfs}. Our analysis in this paper applies to any chargegenesis scenario that populates the entries of the charge vectors on the right-hand side of these equations, except for the lepton flavor asymmetries $Q_{\Delta_\alpha}$.

Strictly speaking, the framework of wash-in leptogenesis is even more general and does not require vanishing initial conditions for the lepton flavor asymmetries. Similarly as in the case of other leptogenesis scenarios in the literature, our analysis may be extended by allowing for the possibility of pre-existing lepton flavor asymmetries; see, e.g., Ref.~\cite{Bertuzzo:2010et}, which discusses the impact of pre-existing asymmetries on flavored thermal leptogenesis. In the strong wash-in regime, the effect of nonvanishing initial conditions, $Q_{\Delta_\alpha} \left(z\ll 1\right) \neq 0$, will be exponentially suppressed, as illustrated by the exact analytical solution in Eq.~(7) in Ref.~\cite{Domcke:2020quw}. Nonvanishing initial conditions in the weak wash-in regime, however, deserve more attention in future work.

Four example solutions of our Boltzmann equations for these initial conditions are shown in Fig.~\ref{fig:BEsols}.
For a given numerical solution, we quantify the efficiency of wash-in leptogenesis, in relation to the optimal outcome mentioned above, $q_{B-L} = -\sfrac{3}{10}\,q_e$, in terms of the efficiency factor
\begin{equation}
\label{eq:kappa}
\kappa = \frac{1}{\sfrac{3}{10}} \left[ Q_{\Delta_e} \left(z\gg 1\right) + Q_{\Delta_\mu} \left(z\gg 1\right) + Q_{\Delta_\tau} \left(z\gg 1\right) \right] \,.
\end{equation}
For given $\kappa = \kappa\left(\widetilde{m}_1,M,p_e,p_\mu\right)$ and primordial input charge $q_e^{\rm ini} \equiv \sfrac{1}{6}\:\bar{\mu}_e^{\rm ini} T^2$, one thus obtains a final baryon asymmetry 
\begin{equation}
\eta_B^0\left(\widetilde{m}_1,M,p_e,p_\mu,q_e^{\rm ini}\right) = C_{\rm sph}\,C_{\rm win}\,\frac{g_{*,s}^0}{g_{*,s}^{\rm SM}}\,\kappa\left.\frac{q_e^{\rm ini}}{n_\gamma}\right|_{\rm ini} \,,
\end{equation}
with sphaleron and wash-in leptogenesis conversion factors $C_{\rm sph} = \sfrac{12}{37}$~\cite{Harvey:1990qw,Laine:1999wv} and $C_{\rm win} = \sfrac{3}{10}$, SM entropy DOFs today and at high temperatures, $g_{*,s}^0 = \sfrac{43}{11}$ and $g_{*,s}^{\rm SM} = \sfrac{427}{4}$, and photon number density $n_\gamma = \zeta(3)\,g_\gamma T^3/\pi^2$, where $g_\gamma = 2$.

The allowed $q_e^{\rm ini}$ range is bounded from above by the requirement that the primordial chiral asymmetry stored in right-handed electrons must not trigger the chiral instability of the SM plasma~\cite{Joyce:1997uy,Boyarsky:2011uy,Akamatsu:2013pjd,Hirono:2015rla,Yamamoto:2016xtu,Rogachevskii:2017uyc,Kamada:2018tcs}. Indeed, $\bar{\mu}_e^{\rm ini}$ (i.e., $q_e^{\rm ini}$ rescaled so that it has units of a chemical potential) must satisfy~\cite{Bodeker:2019ajh}
\begin{equation}
\label{eq:CPIbound}
\left.\frac{\bar{\mu}_e}{T}\right|_{\rm ini} \lesssim 9.6 \times 10^{-4} \,,
\end{equation}
which is equivalent to the condition $\left.\mu_{Y,5}/T\right|_{\rm ini} \lesssim 1.4 \times 10^{-3}$, where $\mu_{Y,5}$ is the chiral chemical potential; before the onset of wash-in leptogenesis, we have $\mu_{Y,5} = \sfrac{711}{481}\,\bar{\mu}_e$~\cite{Domcke:2022kfs,Domcke:2022uue}.
If this upper bound is violated, the SM chiral plasma instability (CPI) will develop and convert the chiral charge stored in right-handed electrons into helical hypermagnetic fields. The helicity stored in these fields will then in turn lead to a drastic overproduction of baryon asymmetry~\cite{Domcke:2022uue} via the mechanism of baryogenesis via helicity decay~\cite{Giovannini:1997eg,Giovannini:1997gp,Giovannini:1999wv,Giovannini:1999by,Bamba:2006km,Bamba:2007hf,Fujita:2016igl,Kamada:2016eeb,Kamada:2016cnb} around the time of the electroweak phase transition.

\begin{figure}
    \centering
    \includegraphics[width=\linewidth]{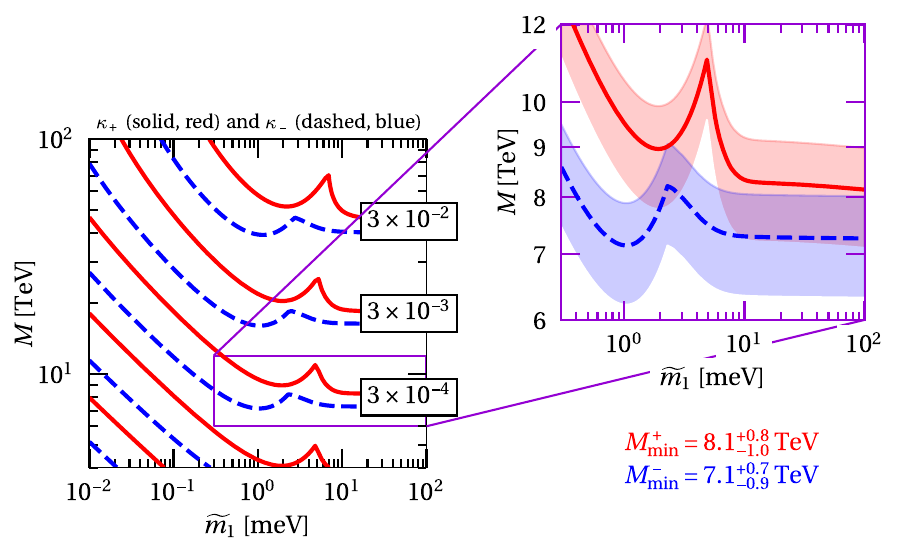}
    \caption{$\kappa_+ > 0$ (solid red contours) and $\kappa_- < 0$ (dashed blue contours) solutions for the efficiency factor, which assume optimal RHN branching ratios $p_{e,\mu,\tau}$, as functions of the neutrino masses $\widetilde{m}_1$ and $M$. The transparent bands in the zoomed-in portion on the right-hand side indicate the ranges $\left|\kappa_\pm\right| = 2\cdots4\times 10^{-4}$, with the central lines corresponding to the $\left|\kappa_\pm\right| = 3\times 10^{-4}$ contours.}
    \label{fig:scan}
\end{figure}

The upper bound on the primordial input charge, together with the requirement of successful baryogenesis, $\eta_B^0 = \eta_B^{\rm obs}$, implies a lower bound on the efficiency factor $\kappa$, 
\begin{equation}
\eta_B^0 \simeq \eta_B^{\rm obs}\,\, \Bigg(\frac{\kappa}{2.6 \times 10^{-4}}\Bigg)\left(\frac{\bar{\mu}_e/T}{9.6 \times 10^{-4}}\right)_{\rm ini} \,.
\end{equation}
In the following, we are therefore interested in identifying the region of parameter space where $\kappa \gtrsim 2.6 \times 10^{-4}$. To this end, we perform a numerical scan over the four-dimensional parameter space spanned by $\widetilde{m}_1$, $M$, $p_e$, and $p_\mu$. More precisely, at each point in the $\widetilde{m}_1$\,--\,$M$ plane, we scan over $p_e \in [0,1]$ and $p_\mu \in [0,1-p_e]$ and identify the branching ratios that either maximize $\kappa$ or $-\kappa$. This analysis provides us with two extremal solutions for the efficiency factor, $\kappa_+ > 0$ and $\kappa_- < 0$, in the $\widetilde{m}_1$\,--\,$M$ plane; see Fig.~\ref{fig:scan}. Both solutions are physically viable, as we can always choose the sign of the initial $q_e^{\rm ini}$ charge appropriately. 

The contour plots of $\kappa_\pm$ in Fig.~\ref{fig:scan} represent the main result of our analysis. A first observation looking at Fig.~\ref{fig:scan} is that the $\kappa_\pm$ contours behave differently at small and large effective neutrino masses, which is reflected in the underlying optimal branching ratios. At $\widetilde{m}_1 \ll 5\,\textrm{meV}$, maximal $\left|\kappa_+\right|$ ($\left|\kappa_-\right|$) is achieved for $p_e = 0$ and $p_{\mu,\tau} = \sfrac{1}{2}$ ($p_e = 1$ and $p_{\mu,\tau} = 0$). Typical solutions of the Boltzmann equations in this regime of parameter space are shown in the top row of Fig.~\ref{fig:BEsols}. At $\widetilde{m}_1 \gg 5\,\textrm{meV}$, on the other hand, $\left|\kappa_+\right|$ ($\left|\kappa_-\right|$) is maximized for $p_e \simeq 1 - 2.0\,\textrm{meV}/\widetilde{m}_1$ ($p_e \simeq 1.1\,\textrm{meV}/\widetilde{m}_1$) in combination with $p_{\mu,\tau} = (1-p_e)/2$ ($p_\mu = 0$ and $p_\tau = 1-p_e$); see the bottom row of Fig.~\ref{fig:BEsols} for typical solutions of the Boltzmann equations.


The $\left|\kappa_\pm\right| = 3 \times 10^{-4}$ contours, furthermore, allow us to estimate lower bounds on the RHN mass $M$. In the case of positive (negative) $\kappa$, optimally chosen branching ratios, an effective neutrino mass around $\widetilde{m}_1 \sim 100\,\textrm{meV}$ ($\widetilde{m}_1 \sim 1\,\textrm{meV}$), and maximally allowed primordial input charge, wash-in leptogenesis can still successfully operate down to RHN masses of around $M \sim 8\,\textrm{TeV}$ ($M \sim 7\,\textrm{TeV}$). In fact, since the upper limit in Eq.~\eqref{eq:CPIbound} is not a perfectly sharp threshold and we rather expect the CPI to become gradually more important around primordial input charges around the threshold value, we estimate the uncertainty of our RHN mass bound by allowing the lower bound on $\kappa$ to vary in the range from $\kappa = 2 \cdots 4 \times 10^{-4}$. In this way, we arrive at lower bounds of
\begin{equation}
\label{eq:Mmin}
M_{\rm min}^+ = 8.1_{-1.0}^{+0.8} \,\textrm{TeV} \,, \qquad M_{\rm min}^- = 7.1_{-0.7}^{+0.9} \,\textrm{TeV} \,,
\end{equation}
where $M_{\rm min}^{\pm}$ refer to the $\kappa_\pm$ solutions in Fig.~\ref{fig:scan}, respectively. 


\medskip\noindent
\textbf{Conclusions}\,---\,The low RHN mass bounds in Eq.~\eqref{eq:Mmin} are a remarkable result, which highlights that, assuming a large primordial input charge near the CPI threshold, the viable RHN mass range for successful wash-in leptogenesis extends around one order of magnitude below the equilibration temperature of the electron Yukawa interaction, $M_{\rm min} \sim 0.1\,T_e$. Our new bound on the RHN mass scale is thus one order of magnitude lower than the naive value of $M_{\rm min} \sim 100\,\textrm{TeV}$ initially stated in Ref.~\cite{Domcke:2020quw} and roughly five orders of magnitude below the classic Davidson--Ibarra bound on the RHN mass in standard thermal leptogenesis. We thus arrive at our main conclusion, which we already anticipated in the introduction: if, in the context of baryogenesis via leptogenesis, RHN interactions only serve the purpose to violate $B\!-\!L$ and $CP$ violation is attributed to new chargegenesis dynamics at higher energies, large viable regions in the neutrino mass parameter space open up; see Fig.~\ref{fig:scan}. The relaxation of the lower bound on the RHN mass scale $M_N$ in wash-in leptogenesis has notably important consequences for model building, as $M_N$ no longer needs to lie several orders of magnitude above the electroweak scale. 


In closing, we remark that the RHN mass bounds in Eq.~\eqref{eq:Mmin} are still subject to remaining theory uncertainties due to our rough treatment of the CPI. In our analysis, we assumed that no CPI will develop as long as the bound in Eq.~\eqref{eq:CPIbound} is satisfied, while baryon number overproduction via helicity decay at the electroweak scale is unavoidable if it is not satisfied. This treatment could be refined by explicitly including the effect of anomalous $q_e$ violation in the $q_e$ Boltzmann equation, in combination with a new transport equation for hypermagnetic helicity (e.g., along the lines of Appendix~C in Ref.~\cite{Domcke:2019mnd}). In general, a proper understanding of the onset of the CPI at large chiral charge $\mu_{Y,5}$, however, calls for dedicated analyses based on magnetohydrodynamical simulations of the chiral plasma at temperatures $T \sim T_e$. We leave such refinements for future work and conclude instead that wash-in leptogenesis provides an intriguing framework for the generation of the BAU that can successfully operate at RHN masses as low as $M_N \sim 7\,\textrm{TeV}$. Wash-in leptogenesis at such a low RHN mass scale requires the presence of a primordial charge asymmetry in right-handed electrons; the exploration of possible chargegenesis mechanisms capable of producing this primordial input charge deserves more attention in future work as well. Finally, it would be interesting to study extended scenarios that allow for the generation of nonzero lepton flavor asymmetries already at the stage of chargegenesis. Scenarios of this type, which would blur the line between the dynamical sectors responsible for lepton flavor and lepton number violation, promise to result in even lower RHN mass bounds and a smooth parametric interpolation between our results in this paper and RHN mass bounds in other leptogenesis scenarios. We encourage more work in this direction.


\medskip\noindent
\textit{Acknowledgments}\,---\,The authors would like to thank Dietrich B\"odeker, Kohei Kamada and Kyohei Mukaida for helpful discussions and comments. M.\,A.\,M.\ acknowledges support from the Deutsche Forschungsgemeinschaft (DFG) Collaborative Research Centre ``Neutrinos and Dark Matter in Astro- and Particle Physics'' (SFB 1258).


\bibliographystyle{JHEP} 
\bibliography{journal_3}


\end{document}